\documentclass{IEEEcsmag}

\usepackage[colorlinks,urlcolor=blue,linkcolor=blue,citecolor=blue]{hyperref}
\expandafter\def\expandafter\UrlBreaks\expandafter{\UrlBreaks\do\/\do\*\do\-\do\~\do\'\do\"\do\-}
\let\labelindent\relax
\usepackage[inline]{enumitem}
\usepackage{upmath,color}
\usepackage{multirow}
\usepackage{ifthen}
\usepackage{booktabs}
\usepackage{graphicx}

\usepackage{xpatch}
\xpatchcmd{\IEEEbiography}
{\fontsize{8.25}{10}\selectfont}
{\normalsize}
{}{}

\newcommand{\TODO}[1]{\textcolor{red}{#1}\GenericWarning{}{LaTeX Warning: TODO: #1}}\newcommand\todo\TODO

\jvol{XX}
\jnum{XX}
\paper{8}
\jmonth{February}
\jname{---||---}
\jtitle{SOBO: A Feedback Bot to Nudge Code Quality in Programming Courses}
\pubyear{2023}

\begin{document}

\sptitle{Software Engineering Education}

\title{SOBO: A Feedback Bot to Nudge Code Quality in Programming Courses}

\author{Sofia Bobadilla}
\affil{KTH Royal Institute of Technology}

\author{Richard Glassey}
\affil{KTH Royal Institute of Technology}

\author{Alexandre Bergel}
\affil{RelationalAI}

\author{Martin Monperrus}
\affil{KTH Royal Institute of Technology}

\markboth{Software Engineering Education}{Software Engineering Education}

\begin{abstract}\looseness-1
Recent research has shown the great potential of automatic feedback in education.
This paper presents SOBO, a bot we designed to automatically provide feedback on code quality to undergraduate students. SOBO has been deployed in a course at the KTH Royal Institute of Technology in Sweden with 130+ students.
Overall, SOBO has analyzed 1687 GitHub repositories and produced 8443 tailored code quality feedback messages to students. The quantitative and qualitative results indicate that SOBO effectively nudges students into adopting code quality best practices without interfering with pedagogical objectives or adding a teaching burden. From this experience, we provide guidelines into how to design and deploy teaching bots in programming courses.
\end{abstract}

\maketitle
 
\section{Introduction}
It is generally accepted that code quality should be taught early and often throughout a computer science (CS) curriculum. Yet, Jansen et al. \cite{jansen2017impact} showed that feedback related to code quality tends to be delivered far too late to have any meaningful impact, typically at the very end of an assignment and beyond any student interest. Furthermore, Östlund et al. \cite{wicklund2022never} demonstrated in a longitudinal study that code quality is not seriously considered by students. However, it is unreasonable to expect busy teachers and teaching assistants to devote significant time delivering targeted feedback on code quality to every student when resources are already stretched.

Whilst tools exist to discover code quality violations, it is  challenging to integrate such professional tools into a student working environment without creating confusion. Our key insight is to encapsulate such a code quality tool into a friendly bot that automates the production of tailored, helpful feedback on code quality. In this paper, we present SOBO, an automatic feedback tool, whose aim is to nudge students towards understanding and applying code quality best practices early in their software engineering journey.

\section{Nudging and Automatic Feedback}

\begin{figure*}
\centerline{\includegraphics[width=30pc]{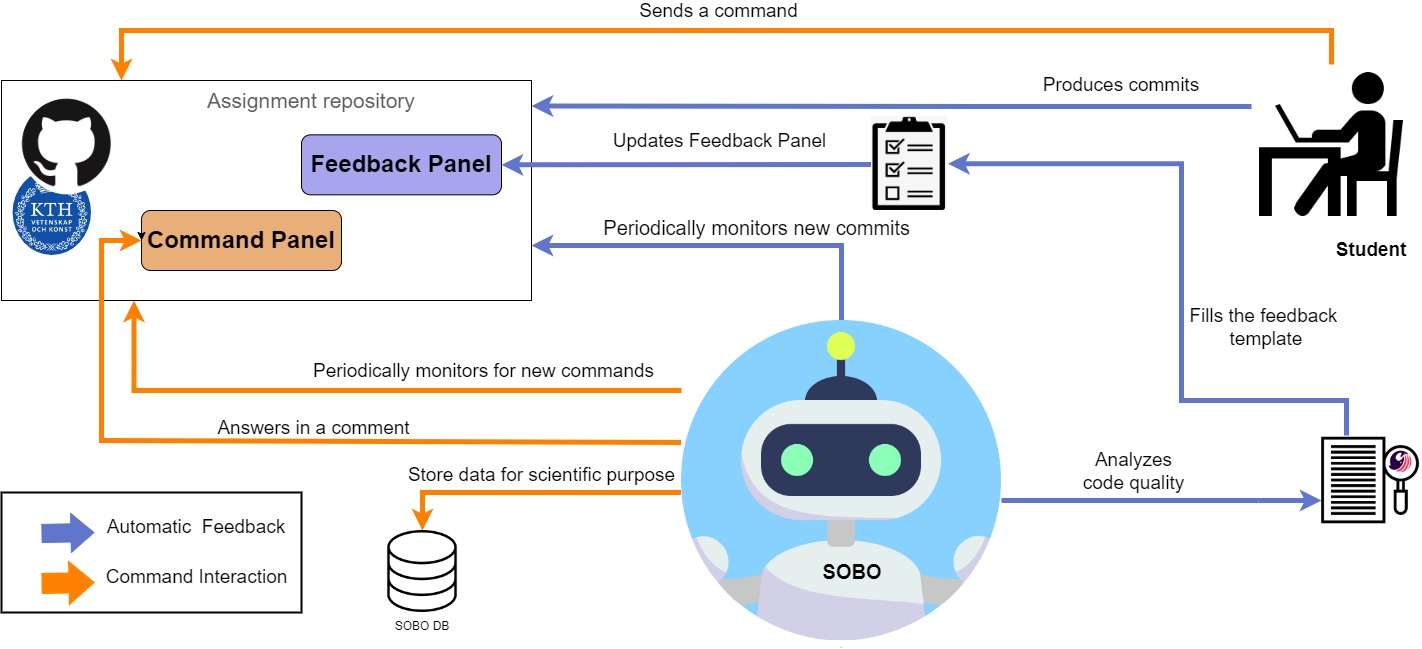}}
\caption{An overview of SOBO, KTH's automatic feedback bot that nudges undergraduate students towards adopting code quality best practices.}\vspace*{-5pt}
\label{workflow}
\end{figure*}

At the highest level, SOBO brings together the ideas of nudging and automation in education. Nudging is a concept that has emerged from behavioral economics. We adopt the definition by Thaler and Sunstein \cite{thaler2009nudge} of nudging being ``any aspect of the choice architecture that alters people's behavior in a predictable way''. In their review on nudging in education, Dammgaard and Nielcen~\cite{DAMGAARD2018313} show that nudge techniques can gently guide students, parents, and teachers toward better educational decisions and attainment. In light of this, we want students to be more aware of code quality - by communicating violations, their meaning and solution - and we hope to alter their behavior towards code quality in a positive way.

Automation in education has become increasingly important to manage ever growing enrollments. Within computer science, introductory courses such as CS1 (programming) and CS2 (algorithms and datastructures) are very popular and as such must find ways to manage large numbers of students, whilst not compromising on quality. For example, Paiva et al.~\cite{10.1145/3513140} finds that manual assessment is a bottleneck task for instructors. As a potential solution, automatic feedback allows for instant and detailed feedback on student attempts without ever overloading the instructor - a rare win-win situation.

In the last decade, several projects related to automatic feedback have been implemented. Studies like Jansen et al.~\cite{jansen2017impact} show that it is a hard task to extract meaningful statistics from experiments related to the usage and impact of automation bots.
Haug, Markus et al.~\cite{Security} address one common challenge among feedback tools which is the fact that the content of the feedback itself can be ignored by the user if the message is not clear enough. To solve the problem, they propose to add code examples of how to fix the error in the feedback. A recent important study~\cite{guidelines}  highlights the importance of clear and concise information in automated feedback. 

Another common challenge among automatic feedback tools is how not to give the direct answer to the task to the students. Zhang et al. \cite{Competitive} and Liu et al. \cite{EndToEnd} have dealt with this problem: they categorize 1) bots that provide feedback directly related to the grades and 2) bots that nudge better techniques but do not affect the student's grades.

Finally, research on code quality in CS1 and CS2 courses highlight two major challenges: (1) the amount of topics to cover, and (2) the assessment of such topics~\cite{jansen2017impact,10.1145/3513140}.  As Jansen et al. observes, most of the educational effort is spent on the aspects of learning the fundamentals of programming, new languages, and development environments. All this research points to 1) automating the code quality feedback and 2) nudging students towards addressing it without any explicit grading demand.
This is what our new bot, called SOBO, does.

\section{SOBO Design}
Motivated by the research in nudging and automation, we next present the design for SOBO by describing the teaching context in which it is deployed, its workflow, the choice of code quality rules, the feedback design, and interaction with the students.

\subsection{Teaching Context}
We design SOBO for a CS1 and CS2 course at KTH Royal Institute of Technology, Stockholm, Sweden. The course is part of the first-year program of the Computer Science (CS) degree and the topics covered include both an introduction to programming in Java (CS1) , as well as an introduction to algorithms and data structures (CS2).

The course's workflow is as follows:
\begin{itemize*}
    \item At the beginning of the year, the students receive instructions about the course stack and they immediately learn about Git/GitHub as a version control and collaboration technology.
    \item Every week, the students receive a unique repository per student group, containing all the relevant task information. They are free to commit/push as often as they like prior to the deadline.
    \item At the end of each week, they have a session with a Teaching Assistant (TA) where they present.
    \item After each session, TAs grade the students' submissions and students have the opportunity to fix issues in case they did not fulfill the requirements. TAs use the GitHub issue tracker to communicate with the students. Per task, the grades can be PASS, FAIL, or KOMPLËTERING when the last one means there are a few mistakes to be corrected.
\end{itemize*}

The course has some features that are relevant to the purpose of this work:
\emph{Consistent course information in one stack.}
The whole course happens on GitHub: students receive repositories for the weekly tasks and commit/push their solutions from the beginning to the end. The GitHub issue tracker is the main channel of communication for the whole course. Should they talk to a bot, the students would naturally do it via an issue tracker.

\emph{Access to historical data.}
The course has been delivered using the same workflow and stack since 2015. Therefore, there is access to data of each previous offering (200 students working in 4000 task repositories over seven years). Previous research has analyzed the data between 2018 and 2022 showing that students are not taking enough action in terms of code quality in their programming assignments \cite{wicklund2022never}.

\subsection{Bot Workflow} 
\begin{figure}
\centerline{\includegraphics[width=18pc]{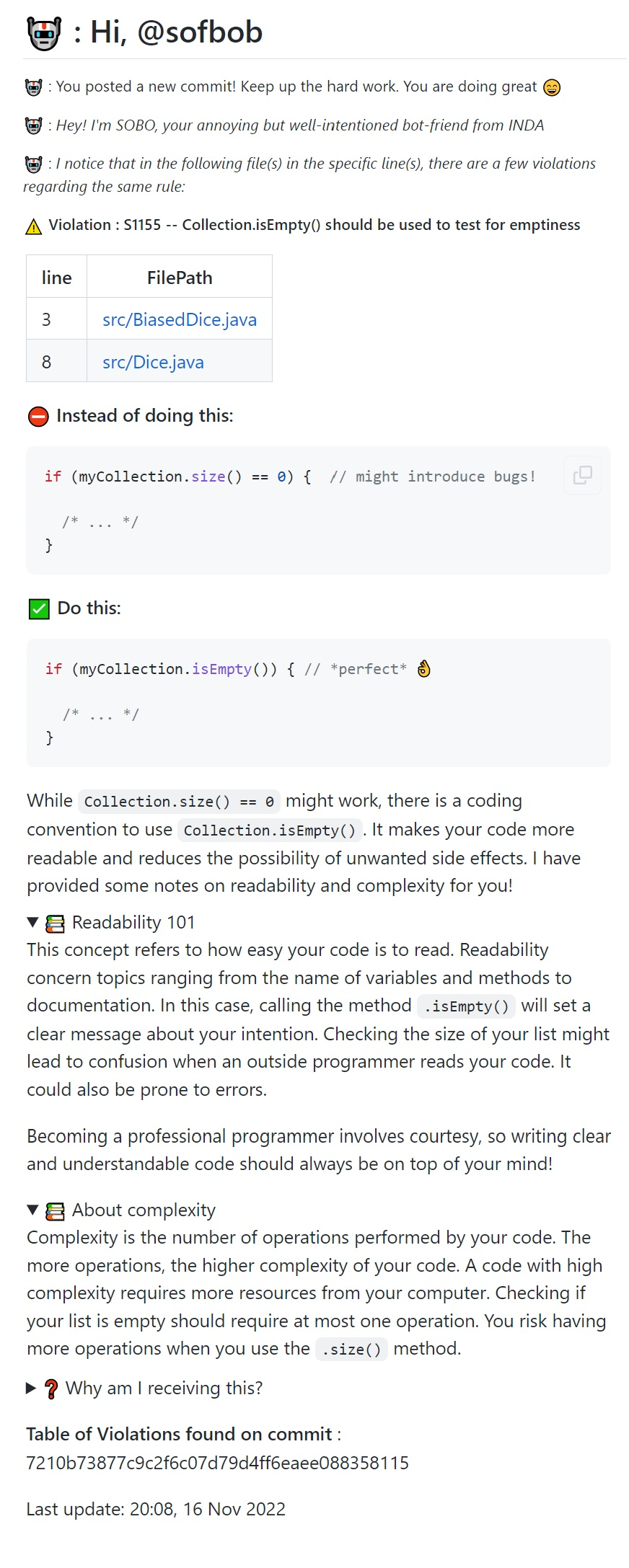}}
\caption{Example of automatic feedback provided by SOBO for code quality rule S1155 by SonarQube~\cite{sonarsource}.}\vspace*{-5pt}
\label{snippet}
\end{figure}

As summarized in Figure~\ref{workflow} SOBO's automatic feedback workflow operates as follows:
\begin{enumerate}
   
    \item When a new programming assignment is announced, a weekly repository is created for each student group and added to SOBO's monitoring list.
    \item Then, SOBO opens a GitHub issue to push feedback messages to come. The first post is a friendly greeting text\footnote{\url{https://github.com/SOBO-bot/templates/blob/main/files/greeting.md}}.
    \item Every time a student produces commits, SOBO performs the following task:
    \begin{enumerate}
        \item It uses a state-of-the-art code quality tool called SonarQube to list the violations in the students' code, incl. the commit hash with the file where the violation is found, the rule violated, and the line of code in the respective file.
        \item Then, it filters each row to confirm the violation comes from a student and not from the provided template~\cite{wicklund2022never}.
        \item It selects the most prevalent violation types from the last commit in order to give feedback on one single code quality aspect (we cannot flood the students, it would be counter-productive).
        \item SOBO collects the template for the selected rule and fills it up with a table indicating the location of the violation(s) and pushes the feedback to the corresponding issue. Figure~\ref{snippet} gives an example of this feedback message.
        \item If there are no violations at all, SOBO posts an encouraging message indicating the submission had no violations.
    \end{enumerate}
\end{enumerate} 

This design is per Wessel et al. \cite{guidelines}, SOBO does not need any configuration from the students.
Also, in case a student does not want to receive feedback, we provide a simple opt-out option to stop receiving automated feedback. 

\subsection{Code Quality Feedback} 
\subsubsection{Violation Reports.} SonarSource for Java \cite{sonarsource} is used to analyze students' code quality. This tool indicates the presence of code quality problems in code.
\subsubsection{Curation of Rules.} To select the relevant code quality rules for students, we categorized all SonarSource rules by using the following criteria:
\begin{enumerate}
    \item Actionable. A clear example of how to fix a problem can be extracted from the rule.
    \item Knowledge gap. The information is understandable by undergraduate students in an introductory programming course.
    \item Actual problem. The prevalence of violations from previous course iterations is used as a relevance indicator (see Östlund, et al. \cite{wicklund2022never} for details).

\end{enumerate}

The curating process was done in collaboration with TAs the course responsible. All rules currently supported by SOBO are described in Table~\ref{tab1}.

\begin{table}
\vspace*{4pt}
\caption{Code quality rules used by SOBO for providing feedback.}
\label{table}
\tablefont
\begin{tabular*}{17.5pc}{@{}p{25pt}p{175pt}<{\raggedright}p{80pt}<{\raggedright}@{}}
\toprule
Rule& Description \\
\colrule
S109  & Magic numbers should not be used  \\[3pt]
S1481 & Unused local variables should be removed \\[3pt]
S1155 & Collection.isEmpty() should be used to test for emptiness \\[3pt]
S1213 & The members of an interface or class declaration should appear in a pre-defined order \\[3pt]
S2119 &  ''Random'' objects should be reused \\[3pt]
\botrule
\end{tabular*}\vspace*{8pt}
\label{tab1}
\end{table}

\subsection{Feedback Design} 

\subsubsection{Relevance of Rules.}
As a nudge tool,  SOBO must filter the information in order to successfully alter the student behavior and it is not perceived as a negative element in the course \cite{thaler2009nudge, guidelines}. Following Guideline~\#2 from Wessel et al.  \cite{guidelines}, the information needs to be centralized to avoid overwhelming the  user. For, SOBO focuses on the most common rule, as the rule that presents the largest amount of violations in a commit, the rule that requires most attention by students. Each automatic feedback provides information related to the most common rule only. 

\subsubsection{Feedback Template.} 
Wessel et al. \cite{guidelines} mentions ``If [feedback] is not actionable, it is not usable''. We make sure that the feedback is clear and self-contained so that there is no need for extra search needed to understand it. Inspired by Paiva et al. \cite{10.1145/3513140}, SOBO messages present knowledge following a template with 1) a presentation of the rule 2) the actual problem in the student's code, 3) and a code example that is actionable. An example is shown in Fig~\ref{snippet}
    
\subsubsection{Frequency of Feedback.}
Students require timely feedback during the process of their assignments \cite{jansen2017impact, Competitive}. Since we want students to have feedback as soon as they push to the repository, SOBO analyzes students' code every 5 seconds. 

\subsubsection{Distribution Channel.}
SOBO gives feedback by posting on a dedicated GitHub issue, named ``SOBO - Commit Analyzer''.
This has two advantages:
\begin{enumerate*}
    \item students always know where to find the analysis of the latest version of the code;
    \item they can activate or deactivate notifications for a particular task by `unsubscribing' to the issue;
\end{enumerate*}

\subsection{Command Language}
Following recommendation \#2 from Wessel et al.~\cite{guidelines}, we bake in SOBO a direct communication channel with the students. A second GitHub issue is created to interact with the teaching bot. In this issue, the students receive a help message with a list of commands to use and interact with SOBO. 
To send a command, the students write a comment on this issue. 
The available commands are: 
    \begin{itemize}
        \item \texttt{<help>}: to get more information about the available commands.
        \item \texttt{<stop>}: to stop all automatic feedback messages on the repository where the command is executed.
        \item \texttt{<go>}: to restart SOBO on the repository where the command is executed.
        \item \texttt{<more commit-id>}: to receive information about a specific commit on the repository where the command is executed.
        \item \texttt{<rule rule-id>}: to get all violations of a specific rule among the ones used in the project on the repository's latest version.
        \item \texttt{<select rule-id>}: to return synthetic data simulating the result of a code injection into the SOBO database.
    \end{itemize}

\subsection{Data Collection}

\emph{Automatic Message.} To keep track of the influence of SOBO, each time a commit is analyzed, a tuple is stored for each violation made by the student.  To identify violations the key of the tuple is the GitHub user, the assignment number, rule, file, line of code, and the commit hash. 

\emph{Command Language.} For the command language, every command sent by the student and resolved by SOBO is stored for further analysis of user interaction. The key pair in this case is the student, the timestamp, and the task.

\subsection{Implementation}
\emph{Technical implementation.}
SOBO is implemented in Java and runs as a daemon on a university server.

\emph{Platform support.}
SOBO assumes that students' repositories are hosted on GitHub, and it supports both public GitHub and enterprise GitHub (KTH uses the latter). SOBO has its own identity, hence its own KTH GitHub account to interact with the students.

\emph{Configuration}
The main configuration of SOBO is a file with a list of student repositories and a file of feedback templates. Per the course workflow, each week, new repositories are added to the bot monitoring list for the new assignments (one assignment = one repository). 
SOBO is publicly available as open-source and can be used by other universities: \url{https://github.com/eclipse/repairnator}.

\section{Experiments \& Results}
SOBO is an end-to-end project, we have deployed it in class at KTH with 130+ undergraduate students.
Below are the results.

\subsection{Preparation}
Before deploying SOBO, we waited for the students to have already been taught about fields, classes, and collections. This means the bot was actually deployed in the 6th week of the course.  The week  the bot started operating, a presentation was made in the classroom and an informative email was sent, to explain the project, how it will benefit and affect them, and how to turn it off in case they do not want to use it.

\subsection{Measuring Effectiveness}
We want to study the number of code quality violations in students' code, for each user, rule, and repository. 
For each repository, we split the number of violations per rule. To track the evolution of each user, we sort the violations by timestamp and compare the amount between two consecutive commits. If the number of violations for the same rule is reduced, then some have been fixed~\cite{sorald}. We assume that the violations have been fixed thanks to SOBO's actions.
In the opposite scenario, if the number increases for the same rule between two commits, we consider that new  violations have been added.

\begin{figure}
\centerline{\includegraphics[width=21pc]{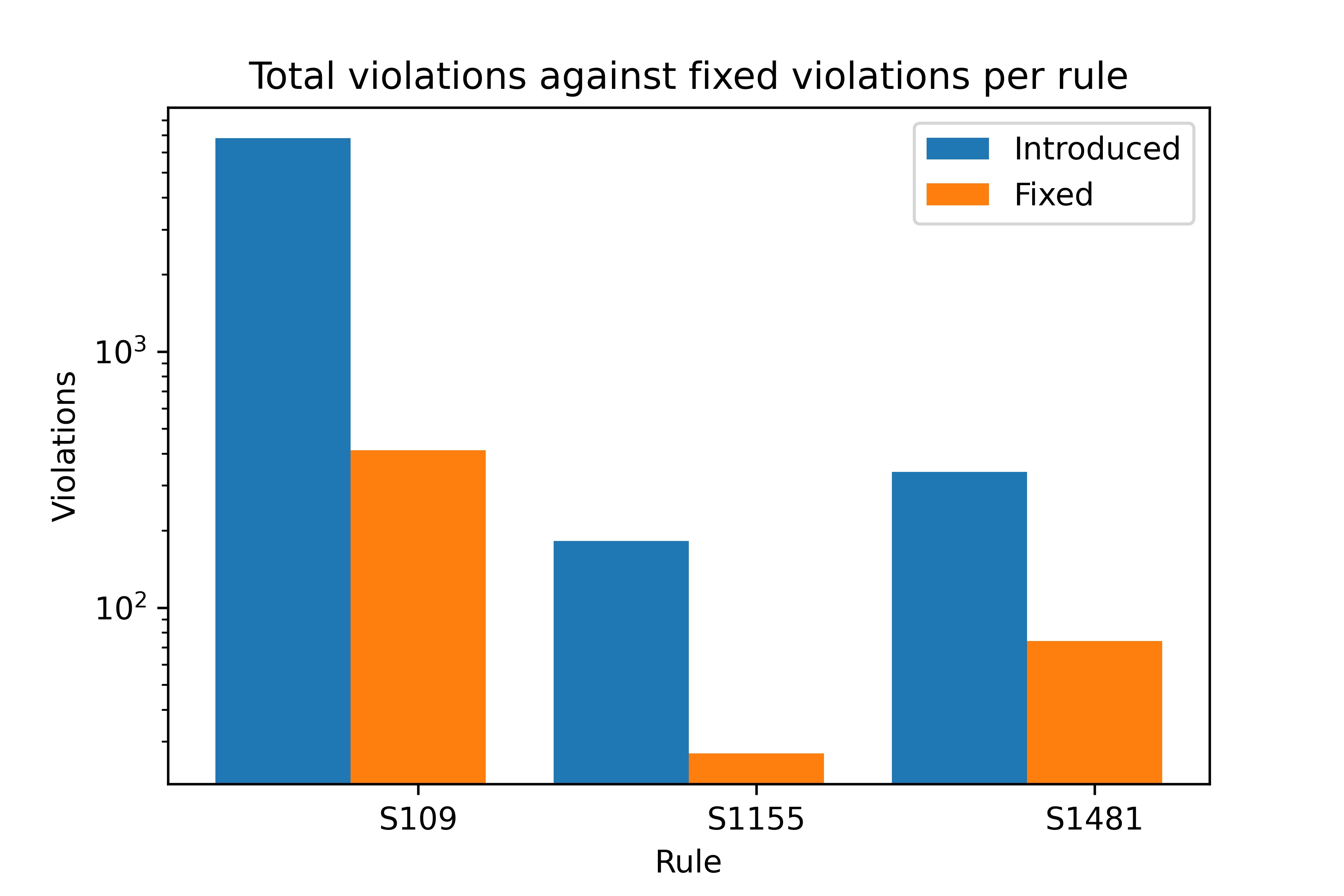}}
\caption{Introduced and fixed violations during the 2022-2023 edition of the course at KTH Royal Institute of Technology for three code quality rules}\vspace*{-5pt}
\label{addFixVio}
\end{figure}

\subsection{Capturing Students' Opinions}

Following the guidelines of Jansen et al.~\cite{jansen2017impact}, we made a survey to study the users' reception of the bot. 
This survey was made and sent 10 weeks after the SOBO integration on the course. The survey was divided into four parts starting with an evaluation of the automatic feedback message, the effectiveness of the provided information, the use of the command language, and the general perception of teaching and/or automatic bots. The survey questions are: \begin{enumerate*}
    \item How clear was the explanation of the rule?
    \item How concise was the explanation of the rule?
    \item How clear was the example of what to do?
    \item How concise was the example of what to do?  
    \item The explanation section helps me to understand how to fix the violation.
    \item The example section helps me to understand how to fix the violation.
    \item I would like to have SOBO integrated on my personal projects.
    \item I would like to have teaching bots in my other courses
\end{enumerate*}

\begin{figure}
\centerline{\includegraphics[width=21pc]{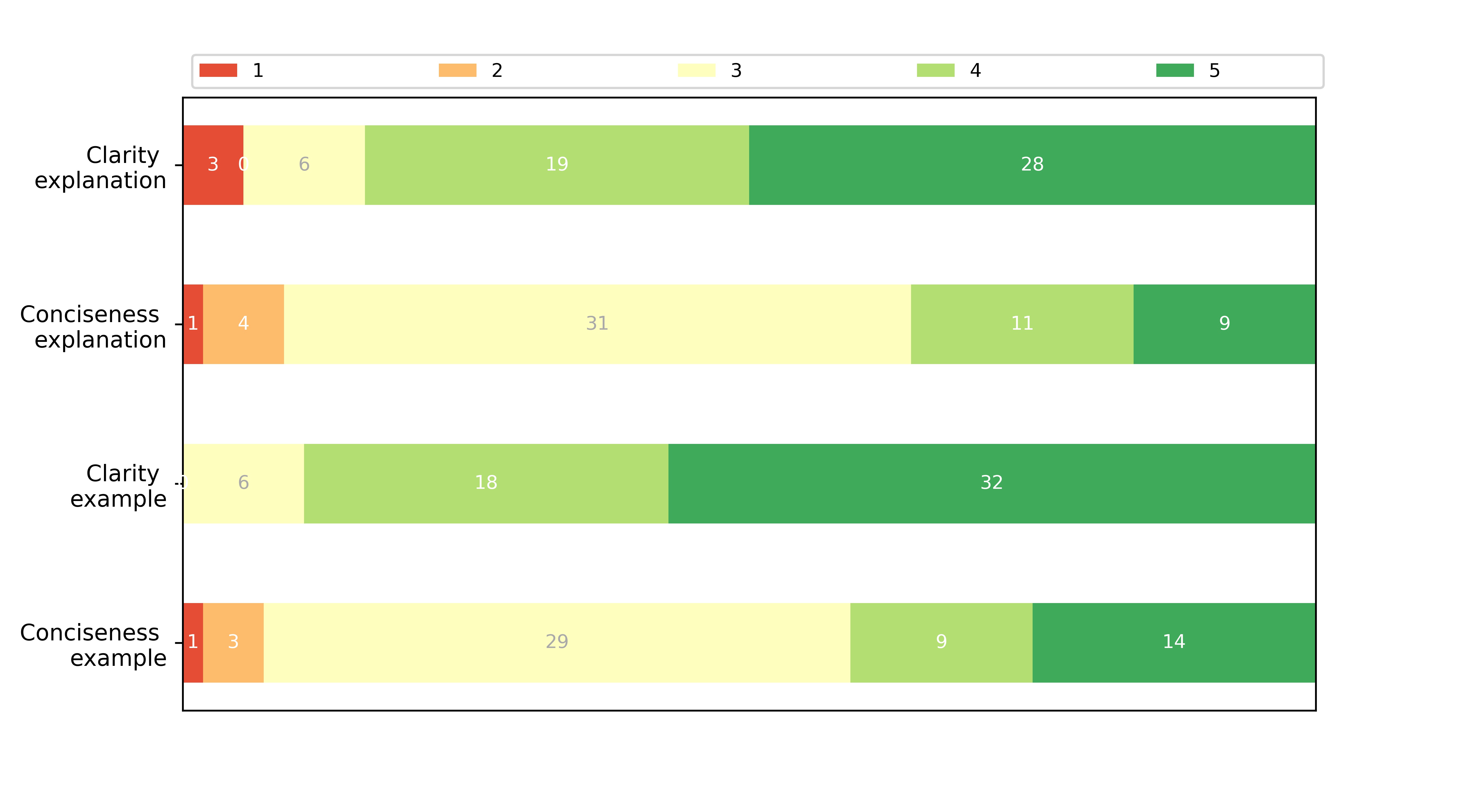}}
\centerline{\includegraphics[width=21pc]{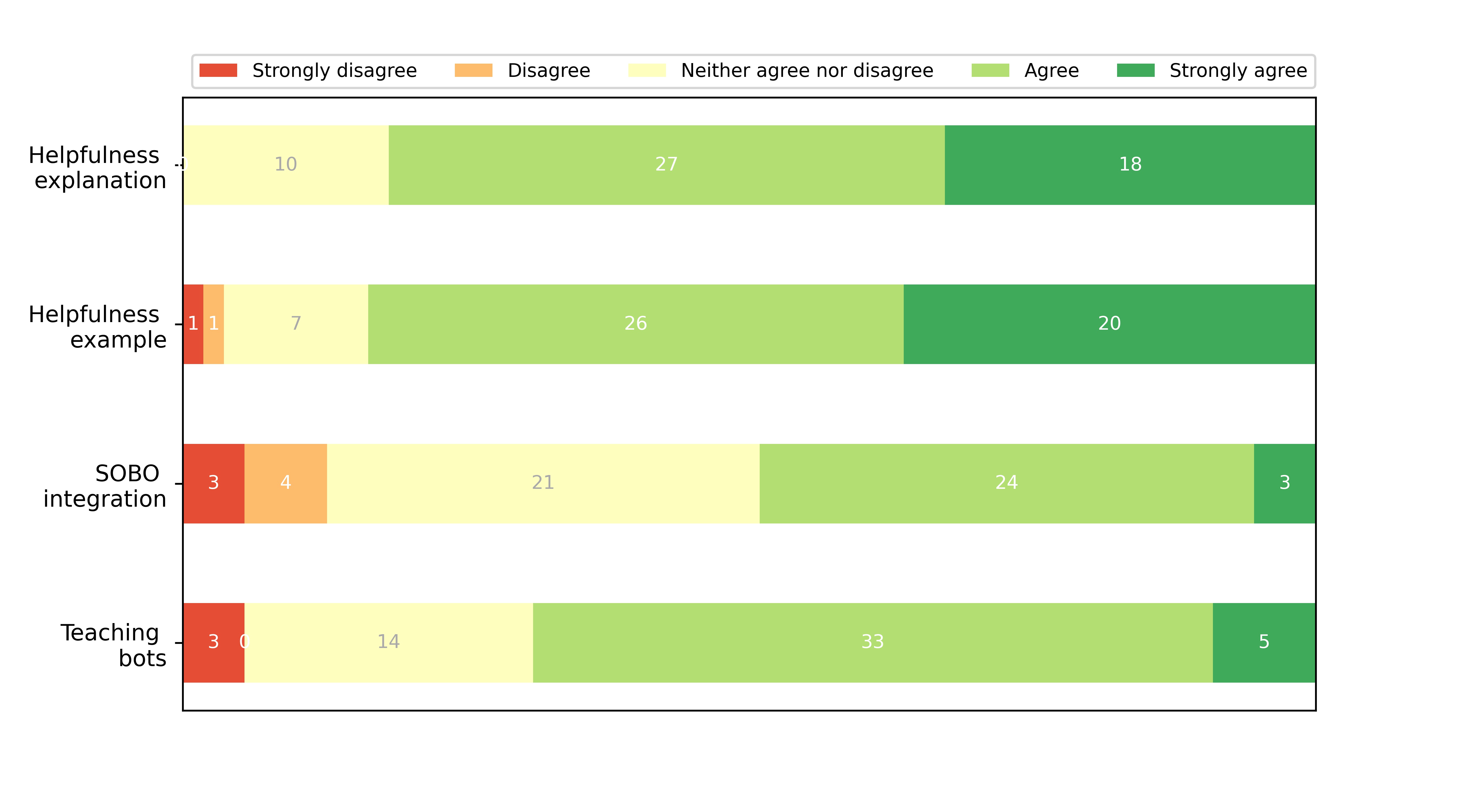}}
\caption{Survey results: students' evaluation on the automatic feedback in terms of clarity, conciseness, and usefulness. }\vspace*{-5pt}
\label{surveyImage}
\end{figure}

\subsection{Results}

\emph{Quantitative. }
Figure~\ref{addFixVio} shows the number of added and removed code quality violations overall student groups for the three most prevalent rules (S109, S1155, S1481).
On the negative side, students add more violations than they remove.
This means that the nudging is not perfectly effective, which is as expected because it does not take part in the mandatory graded objectives.
On the positive side, we do see 513 code quality violations fix by students.
We cannot affirm that this is causally related to SOBO's feedback messages but we have good reasons to believe so. 
First, there are 19 commits messages showing mentioning that the changes were made thanks to SOBO.
Second, the qualitative survey provides solid evidence.

\emph{Qualitative. }
From a total of 131 students, we got 57 answers, reaching 43\% of the users.
The results of the survey are presented in Figure~\ref{surveyImage}.
The student's responses on the survey show that the automatic feedback was clear (Q1, Q3) and contained enough information for them to fix the violation (Q5, Q6).  
The survey also shows that the template is probably too verbose (Q2, Q4). A future iteration of the course will fix that aspect.

In the free comments, students mentioned a lot of messages were related to rule S109 (``Magic Number'').
We believe that this is due to the fact that it is the rule SOBO reports the most.
For this rule in particular, students found the ``Instead of doing this'' part of the template to be very explicit, and that the whole example section allows them to: ``learn how to make the code better''.

\subsection{Impact of the Bot on Pedagogical Objectives}

The bot did not intrude in the main course workflow. Whilst some initial classroom time was used to explain the presence and purpose of SOBO to students, little further effort was required. Neither the course responsible nor teaching assistants reported extra negative burdens related to SOBO and could focus on the normal delivery of the course content. To that extent, the code quality aspect of the course was entirely delegated to the SOBO bot. Consequently, we envision expanding its scope to more aspects of code quality and supporting more pedagogically relevant rules.

\section{Guidelines for Bots in Education}
Based on our experience, here are important guidelines for educators who would like to integrate nudging bots into their courses.

\textbf{Focus on the Message:}
Previous work has mentioned that the key to automatic feedback in education lies in the feedback message that will be delivered~\cite{jansen2017impact,10.1145/3513140}. It is essential not to overestimate the knowledge of the students. SOBO messages were designed to be self-contained and present all the information needed in order to solve the code quality problem. Iterate over the feedback design with focus groups gathering both professors, teaching assistants, and students.

\textbf{Delivering the Message:}
When selecting the distribution channel for the feedback, it is essential to be very clear with the students about the location of the feedback: which platform? which link? In our case, we had one single channel: a specific SOBO GitHub issue, one per student repository.
Avoid changes of applications and channels to deliver automated feedback. 

\textbf{Respect the Context:}
We suggest deploying a teaching bot with careful consideration of the teaching context.
This concerns the workflow of the course(s): when is the bot introduced? for which objectives?
This also concerns the platform: it is better to be integrated into the platform already being used than introduce an additional one. Spend dedicated time explaining the bot's intention and scope in the classroom.

\textbf{Lower the Configuration Barrier:}
Ideally, SOBO can be reused by any other course with only a small configuration at the beginning of the course. 
Extremely custom tools are hard to deploy in new environments, whereas good configurability improves the scalability of automated feedback tools \cite{Artemis,10.1145/3513140}. Adopt software engineering reuse best practices when implementing your own nudging bot.

\section{Conclusion \& Future Work}
It has been shown that there is a lack of feedback related to code quality in CS1 \& CS2 courses. Leaving code quality unaddressed simply delays the cost, which manifests itself as technical debt in future projects. In this paper, we have presented SOBO, a nudging bot that helps students to understand and adopt good code quality practices by generating feedback at the right place and time for maximum impact.

In future iterations, SOBO will track previous student submissions in order to personalize the provided feedback. By tracking previous attempts, the feedback can mention specific ways to help the problem faced by the student.
Also, since not all CS1 \& CS2 courses are taught using Java, an interesting community effort is to extend SOBO to other programming languages (eg. Python), increasing the impact and relevance of the whole nudging bot project. 

\def\refname{References}
\bibliography{main}{}
\bibliographystyle{plain}

\newpage

\begin{IEEEbiography}{\includegraphics[width=1.15in]{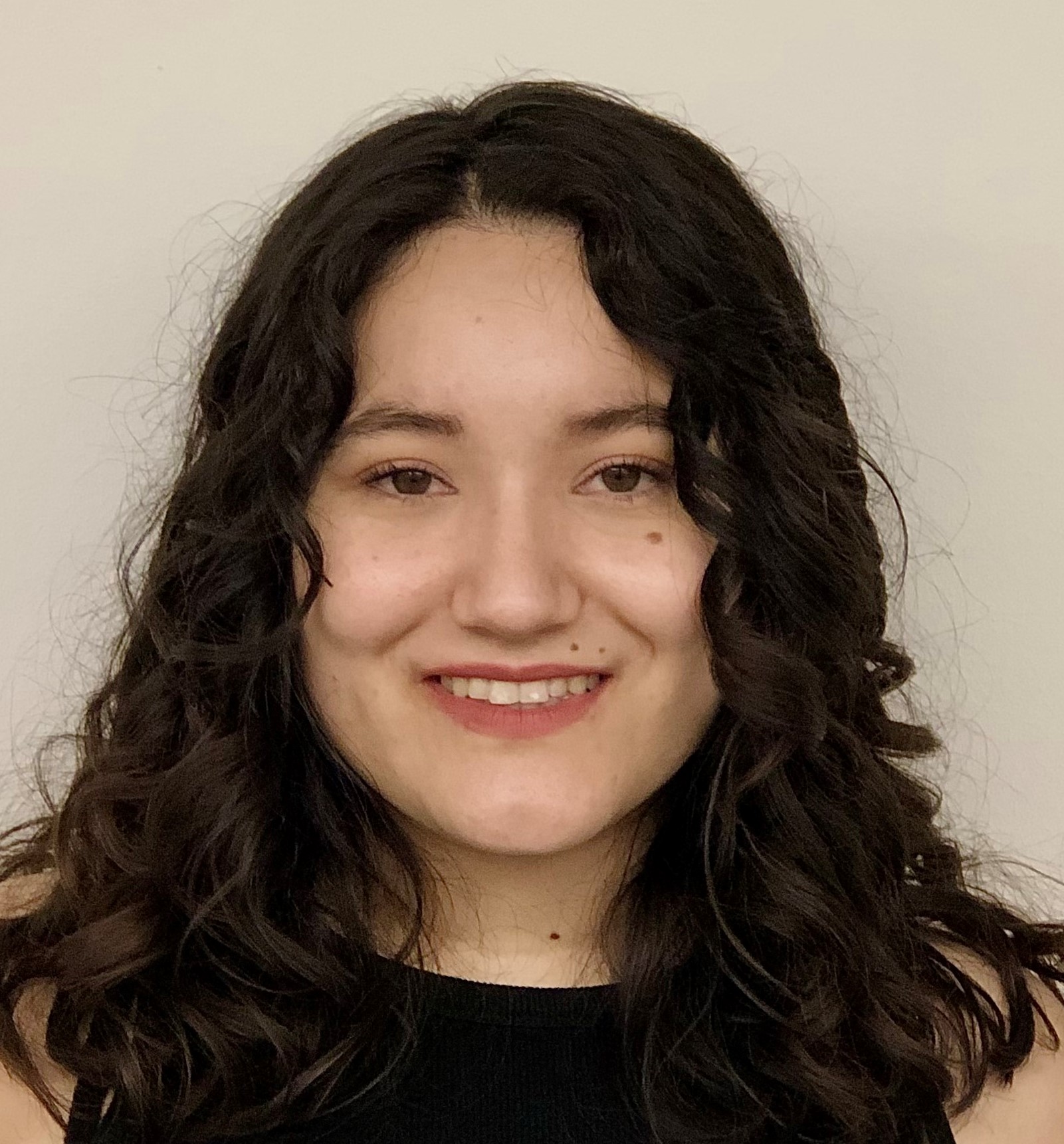}} Sofia Bobadilla is a research engineer at KTH Royal Institute of Technology, Stockholm, Sweden. Her research interest includes bots in software development, software reliability, data science, and multimedia information retrieval. Bobadilla obtained her bachelor's degree in Computer Science from the University of Chile, Santiago, Chile. Contact her at sofbob@kth.se.
\end{IEEEbiography}

\begin{IEEEbiography}{\includegraphics[width=1.15in]{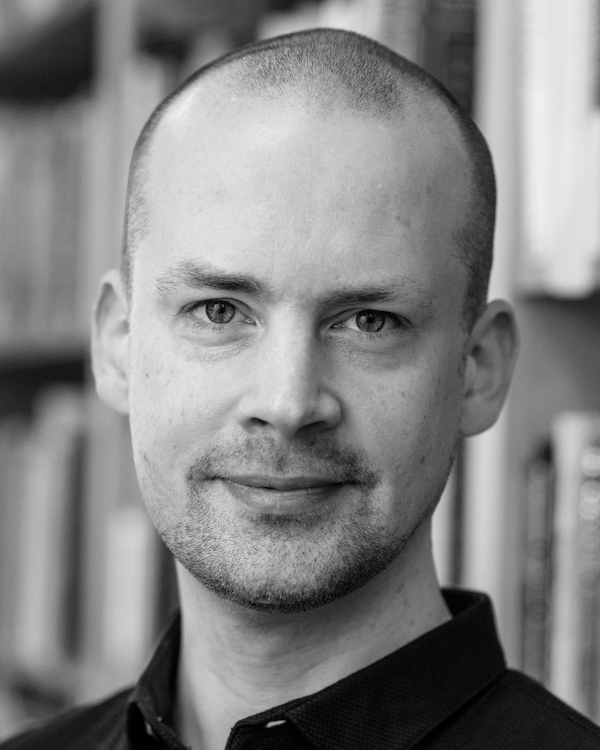}} Richard Glassey
is a teacher at KTH Royal Institute of Technology, Stockholm, Sweden.  His current research interests include computer science education, learning analytics, and sustainable education. Glassey received his Ph.D. in computer science from the University of Strathclyde, Glasgow, Scotland. He is a member of ACM. Contact him at glassey@kth.se.\vspace*{8pt}
\end{IEEEbiography}

\begin{IEEEbiography}{\includegraphics[width=1.15in]{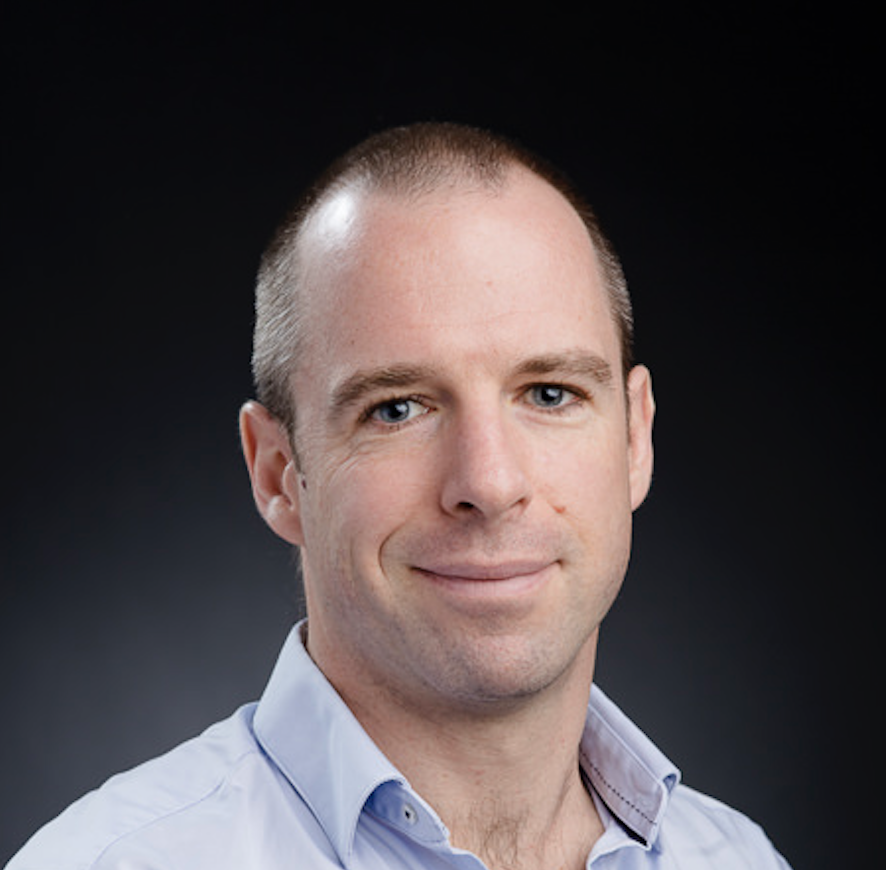}} Alexandre Bergel {\,} is a computer scientist at RelationalAI, Switzerland. His current researches are on designing tools and methodologies to improve the overall performance and internal quality of software systems and databases by employing profiling, visualization, and artificial intelligence techniques. Contact him at alexandre.bergel@me.com.\vspace*{8pt}
\end{IEEEbiography}

\begin{IEEEbiography}{\includegraphics[width=1.15in]{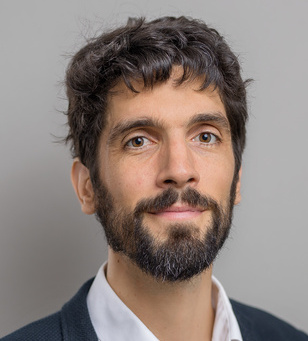}} {\,} Martin Monperrus is Professor of Software Technology at KTH Royal Institute of Technology.  His research lies in the field of software engineering with a current focus on automatic program repair, AI on code and program hardening. He received a Ph.D. from the University of Rennes, and a Master's degree from Compiègne University of Technology.  Contact him at monperrus@kth.se. 
\end{IEEEbiography}

\newpage

\end{document}